\documentclass[12pt]{iopart}
\usepackage{amssymb}
\usepackage{epsfig}
\usepackage{graphicx}

\begin{document}

\title{Linear Response for Confined Particles}
\author{Jeffrey Wrighton$^1$, James Dufty$^1$, Christian Henning$^2$ and Michael Bonitz$^2$}

\begin{abstract}
The dynamics of fluctuations is considered for electrons near a
positive ion or for charges in a confining trap. The stationary
nonuniform equilibrium densities are discussed and contrasted. The
linear response function for small perturbations of this nonuniform
state is calculated from a linear Markov kinetic theory whose
generator for the dynamics is exact in the short time limit. The
kinetic equation is solved in terms of an effective mean field
single particle dynamics determined by the local density and
dynamical screening by a dielectric function for the non-uniform
system. The autocorrelation function for the total force on the
charges is discussed.
\end{abstract}

\pacs{52.27.-h, 52.25.-b, 51.10.+y, 52.58c}

\address{$^1$Department of Physics, University of Florida, Gainesville,
FL 32611}
\address{$^2$Institut f\"{u}r Theoretische Physik und Astrophysik, Christian-Albrechts-Universit\"{a}t zu Kiel, D-24098 Kiel, Germany}
\ead{dufty@phys.ufl.edu}


\section{Introduction}

\label{sec1}

Consider a system of $N$ charges in a spherical container of radius $R$. An
external potential centered at the origin exerts an attractive central
force. The Hamiltonian is
\begin{equation}
H=\sum_{\alpha =1}^{N}\left( \frac{1}{2}mv_{\alpha }^{2}+V_{c}\left(
r_{\alpha }\right) +V_{b}\left( r_{\alpha }\right) +V_{w}\left( r_{\alpha
}\right) \right) +\frac{1}{2}\sum_{\alpha }^{N}\sum_{\gamma \neq \alpha
}^{N}V(r_{\alpha \gamma }).  \label{2.1}
\end{equation}%
where $\mathbf{r}_{\alpha }$ and $\mathbf{v}_{\alpha }$ are the position and
velocity of charge $\alpha $. The repulsive interaction potential between
particles $\alpha $ and $\gamma $ is denoted by $V(r_{\alpha \gamma })$
where $r_{\alpha \gamma }\equiv \left\vert \mathbf{r}_{\alpha }-\mathbf{r}%
_{\gamma }\right\vert $. The external ``confinement" potential
$V_{c}\left( r_{\alpha }\right) $ is the same for all particles, and
$V_{w}\left( r_{\alpha }\right) $ is the wall potential that is zero
inside the container and infinite otherwise. Charged systems are of
direct relevance e.g. to dusty plasmas and ions in traps. For
neutral systems $V_{b}\left( r_{\alpha }\right) $ is the interaction
of each particle with a uniform neutralizing background,
corresponding to an OCP or jellium with an attractive trap at the
origin.

An example that has been studied recently is jellium with a point
positive ion of charge number $Z$ at the origin
\cite{Talin,Wrighton}. At equilibrium the electron density is
enhanced (partial confinement) near the origin and approaches a
uniform limit far from the ion, for sufficiently large $R$, due to
charge neutrality. An opposite extreme is a system of charges in a
strong trap such that they are localized in a finite domain away
from the wall and their density vanishes outside this domain
(complete confinement). There is a continuous mapping between these
two limiting cases, controlled by the relative strengths of the
repulsion between particles and the strength and form of the
confining potential. The discussion of these cases here will be
limited to classical statistical mechanics.

\section{Equilibrium density}

In the absence of a confining potential the equilibrium density may be
uniform (charge neutral) or nonuniform (charged). In general the confining
potential will induce a nonuniform equilibrium density in any case. Formally
it is determined from the exact Yvon-Born-Green equation \cite{Hansen}%
\begin{equation}
\frac{dn\left( r_{1}\right) }{dr_{1}}+\beta n\left( r_{1}\right) \frac{%
dV_{c}\left( r_{1}\right) }{dr_{1}}=-\beta \int d\mathbf{r}_{2}\left(
n\left( \mathbf{r}_{1},\mathbf{r}_{2}\right) -n\left( \mathbf{r}_{1}\right)
n_{b}\right) \frac{dV\left( r_{21}\right) }{dr_{1}},  \label{3.1}
\end{equation}%
where $n\left( \mathbf{r}_{1},\mathbf{r}_{2}\right) $ is the joint density
for two charges and $n_{b}$ is the density of the uniform neutralizing
background. A reasonable approximation is the hypernetted chain
approximation (HNC) \cite{Hansen,Talin1},%
\begin{equation}
\frac{d}{dr_{1}}\left[ \ln n\left( r_{1}\right) +\beta V_{c}\left(
r_{1}\right) -\int d\mathbf{r}_{2}\left( n\left( r_{2}\right) -n_{b}\right)
c\left( r_{21}\right) \right] =0.  \label{3.3}
\end{equation}

The solutions to this equation are quite rich, reflecting the
competition between the attraction of $V_{c}$ and the renormalized
repulsion represented by $c\left( r_{21}\right) $. They have been
described in some detail for the case of the positive ion in jellium
\cite{Talin1}. Here we report on some preliminary results for the
qualitatively different case of weakly coupled charges $q$ in a
harmonic trap $V_{c}\left( r\right) =kr^{2}/2$ with $n_{b}=0$. There
is then a competition between Coulomb repulsion and attraction by
the central trap, with charge density enhanced near the wall or near
the center, respectively. These effects are exactly balanced when
$\omega _{m}=\omega _{c}$, leading to a uniform density at all
temperatures. Here $\omega _{m}=\omega _{p}/\sqrt{3}$ is the Mie
plasma frequency ($\omega _{p}^{2}=4\pi nq^{2}/m$) and $\omega
_{c}=\sqrt{k/m}$ is the frequency associated with the harmonic trap.
More generally, when $\omega _{c}>\omega _{m}$ the charges are
increasingly drawn away from the wall as the temperature is lowered,
approaching at $T=0$ a uniform density for $r<R_{0}=\left( \omega
_{m}/\omega _{c}\right) ^{2/3}R$ and zero density for $R_{0}<r<R$
\cite{Henning}. In the opposite case of $\omega _{c}<\omega _{m}$
the Coulomb repulsion dominates and the particle density is enhanced
at the wall. As the temperature is lowered the density becomes sharp
at the walls - a ``Coulomb explosion" that is restrained by the
external walls. This behavior is
illustrated in Fig. 1a) for $\omega _{c}/\omega _{m}=2$ and in Fig. 1b) for $%
\omega _{c}/\omega _{m}=1/2$, at $T=1$, $0.1$, and $0.01$.

\begin{figure}[t]
 \includegraphics{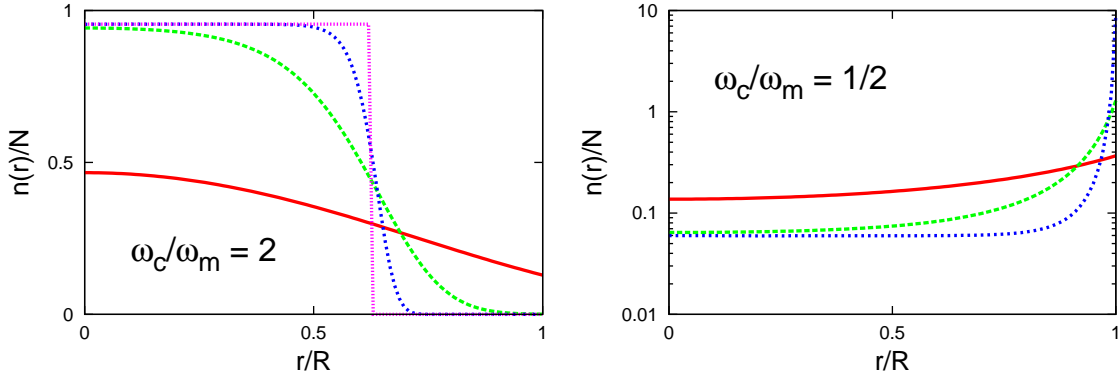}
\caption{Temperature dependence of densities for trap-dominated
(left) and Coulomb-dominated (right) conditions.} \label{fig1}
\end{figure}

\section{Linear Response}

Consider a perturbation of the nonuniform equilibrium state by an
external potential of the form $U_{ext}\left( t\right) =\sum_{\alpha
=1}^{N}V_{ext}\left( r_{\alpha },t\right) =\int
d\mathbf{r}V_{ext}\left( r,t\right) \widehat{n}\left(
\mathbf{r}\right) $, where $\widehat{n}\left( \mathbf{r}\right) $ is
the phase function representing the particle density. Then, to
linear order in $V_{ext}\left( r,t\right) $ the response of the
average particle density to this perturbation is \cite{McLennan}
\begin{equation}
\delta \left\langle \widehat{n}(\mathbf{r,}t)\right\rangle
=\int_{0}^{t}dt^{\prime }\int d\mathbf{r}^{\prime }\chi \left( \mathbf{r},%
\mathbf{r}^{\prime },t-t^{\prime }\right) V_{ext}\left( r^{\prime
},t^{\prime }\right) ,  \label{4.2}
\end{equation}%
with the response function%
\begin{equation}
\chi \left( \mathbf{r},\mathbf{r}^{\prime },t\right) =-\beta \partial
_{t}\left\langle \widehat{n}\left( \mathbf{r}\right) \widehat{n}\left(
\mathbf{r}^{\prime },-t\right) \right\rangle _{e}=\beta \nabla _{\mathbf{r}%
^{\prime }}\cdot \left\langle \widehat{n}\left( \mathbf{r},t\right) \mathbf{j%
}\left( \mathbf{r}^{\prime }\right) \right\rangle _{e}.  \label{4.3}
\end{equation}
The brackets $\left\langle {}\right\rangle _{e}$ denote an equilibrium
ensemble average. The second equality follows from the conservation law $%
\partial _{t}\widehat{n}\left( \mathbf{r},t\right) +\nabla _{\mathbf{r}%
}\cdot \mathbf{j}\left( \mathbf{r},t\right) =0$, and the stationarity of the
equilibrium ensemble. The response function $\chi \left( \mathbf{r},\mathbf{r%
}^{\prime },t\right) $ provides an appropriate object for studying the
possible modes of excitation that are supported by the system. An
alternative equivalent instrument is the dielectric function  $%
\epsilon \left( \mathbf{r},\mathbf{r}^{\prime };t\right) $ defined in analogy with electrodynamics%
\begin{equation}
\epsilon ^{-1}\left( \mathbf{r},\mathbf{r}^{\prime };t\right) \equiv \delta
\left( \mathbf{r}-\mathbf{r}^{\prime }\right) \delta \left( t\right) +\int d%
\mathbf{r}^{\prime \prime }\chi \left( \mathbf{r},\mathbf{r}^{\prime \prime
};t\right) V(\left\vert \mathbf{r}^{\prime \prime }-\mathbf{r}^{\prime
}\right\vert ),  \label{4.4}
\end{equation}%

\section{Markov Kinetic Theory}

The evaluation of the response function is a difficult many-body
problem. The conditions of interest include both strong confinement
and possibly strong charge correlations, so there is no small
parameter available for simplifications. Instead, a non perturbative
Markov kinetic theory has been discussed recently in this context
\cite{Wrighton}. It is based on approximating the formal generator
for dynamics in the single particle phase space by its exact form at
$t=0$. This leads to a mean field theory of the linear Vlasov form,
but with both the confining potential and the charge - charge
potential renormalized by the initial equilibrium correlations. The
analysis of reference \cite{Wrighton} can be extended in a
straightforward
way to the system considered here with the result%
\begin{equation}
\chi \left( \mathbf{r},\mathbf{r}^{\prime },t\right) \rightarrow
\int_{0}^{t}dt^{\prime }\int d\mathbf{r}_{1}\overline{\epsilon }^{-1}\left(
\mathbf{r},\mathbf{r}_{1};t-t^{\prime }\right) \chi _{0}\left( \mathbf{r}%
_{1},\mathbf{r}^{\prime },t^{\prime }\right) .  \label{5.1}
\end{equation}%
where $\chi _{0}\left( \mathbf{r}^{\prime \prime },\mathbf{r}^{\prime
},z\right) $ is the response function for confined particles without the
interparticle interactions ($V(\left\vert \mathbf{r}-\mathbf{r}^{\prime
}\right\vert )=0$)
\begin{equation}
\chi _{0}\left( \mathbf{r},\mathbf{r}^{\prime },t\right) =-\beta n\left(
\mathbf{r}\right) \int d\mathbf{v}\phi \left( v\right) e^{-\mathcal{L}_{0}t}%
\mathbf{v}\cdot \nabla _{\mathbf{r}}\delta \left( \mathbf{r}-\mathbf{r}%
^{\prime }\right) ,  \label{5.2}
\end{equation}%
$\phi \left( v\right)$ is the Maxwellian, and where the generator
$\mathcal{L}_{0}$ for the effective single particle
dynamics is%
\begin{equation}
\mathcal{L}_{0}=\mathbf{v}\cdot \nabla _{\mathbf{r}}-m^{-1}\nabla _{\mathbf{r%
}}\mathcal{V}_{c}\left( r\right) \cdot \nabla _{\mathbf{v}}.  \label{5.3}
\end{equation}%
The renormalized confinement potential is determined from the equilibrium
density by $\mathcal{V}_{c}\left( r\right) \equiv -\beta ^{-1}\ln n\left(
r\right) $. Also, $\overline{\epsilon }\left( \mathbf{r},\mathbf{r}^{\prime
};t\right) $ is
\begin{equation}
\overline{\epsilon }\left( \mathbf{r},\mathbf{r}^{\prime };t\right) =\delta
\left( \mathbf{r}-\mathbf{r}^{\prime }\right) \delta \left( t\right) -\int d%
\mathbf{r}^{\prime \prime }\chi _{0}\left( \mathbf{r},\mathbf{r}^{\prime
\prime },t\right) \mathcal{V}(\mathbf{r}^{\prime \prime },\mathbf{r}^{\prime
}),  \label{5.4}
\end{equation}%
where the renormalized charge - charge potential, $\mathcal{V}\left( \mathbf{%
r},\mathbf{r}^{\prime }\right) $, is defined in terms of the direct
correlation function by $\mathcal{V}\left( \mathbf{r},\mathbf{r}^{\prime
}\right) =-\beta ^{-1}c\left( \mathbf{r},\mathbf{r}^{\prime }\right) $.

Further progress requires evaluation of $\chi _{0}\left( \mathbf{r},\mathbf{r%
}^{\prime },t\right) $ whose dynamics is determined from the effective
single particle charge dynamics in the presence of the confinement
potential. This can be quite complex over the whole range of weak to strong
confinement, and has been discussed in some detail for the case of a central
positive ion in jellium \cite{Wrighton}. The autocorrelation function for
the total force on the charges was studied as a function of the charge
number on the central ion. It was found that a simple representation of the
single particle trajectories in terms of bound and free states provided an
accurate analytical description from weak to strong ion - electron
interaction.

Preliminary studies of the harmonic trap suggest the possibility of a
simplification in that case as well. As noted above, the low temperature
limit leads to a uniform density inside a sphere of radius $R_{0}$. Since
the effective potential $\mathcal{V}_{c}\left( r\right) $ for the dynamics
is determined from this potential, $\chi _{0}\left( \mathbf{r},\mathbf{r}%
^{\prime },t\right) $ becomes the correlation function for free
particles inside a sphere. An exact evaluation of the force
autocorrelation function in terms of all response frequencies is
possible in this case \cite{Marchetti}.

\section{Acknowledgement}

This research was supported by the NSF/DOE Partnership in Basic
Plasma Science and Engineering under the Department of Energy award
DE-FG02-07ER54946.

\end{document}